\documentclass[lettersize,journal]{IEEEtran}
\IEEEoverridecommandlockouts
\usepackage{lineno}
\usepackage{cite,url}
\usepackage{hyperref}
\usepackage{amsmath,amssymb,amsfonts}
\usepackage{amsthm}

\usepackage{graphicx}
\usepackage{textcomp}
\usepackage{xcolor}
\usepackage{graphicx}
\usepackage{float}
\usepackage{textcomp}
\usepackage{stfloats}
\usepackage{amsmath}
\usepackage{amsfonts,amssymb}
\usepackage{mathrsfs}
\usepackage{mathtools}
\usepackage{multirow}
\usepackage{array}
\usepackage{url}
\usepackage{fancyhdr}
\usepackage{mdwmath}
\usepackage{mdwtab}
\usepackage{setspace}
\usepackage{amsmath,amsfonts}
\usepackage{algorithmic}
\usepackage{algorithm}
\usepackage{array}
\usepackage[caption=false,font=normalsize,labelfont=sf,textfont=sf]{subfig}
\usepackage{textcomp}
\usepackage{stfloats}
\usepackage{url,hyperref}
\usepackage{verbatim}
\usepackage{graphicx}
\usepackage{cite}
\usepackage{amsfonts,amssymb}
\usepackage{mathrsfs}
\usepackage{mathtools}
\usepackage{bm}
\usepackage{multirow}
\usepackage{amsthm,bbm,color}
\newtheorem{remark}{Remark}
\theoremstyle{definition}
\newtheorem{theorem}{Theorem}

\newtheorem{lemma}{Lemma}

\newtheorem{corollary}{Corollary}

\makeatletter
\begin{document}
\title{Array Gain for Pinching-Antenna Systems (PASS)}
\author{Chongjun~Ouyang, Zhaolin~Wang, Yuanwei~Liu, and Zhiguo~Ding
\thanks{C. Ouyang and Z. Wang are with the School of Electronic Engineering and Computer Science, Queen Mary University of London, London E1 4NS, U.K. (e-mail: \{c.ouyang, zhaolin.wang\}@qmul.ac.uk).}
\thanks{Y. Liu is with the Department of Electrical and Electronic Engineering, The University of Hong Kong, Hong Kong (email: yuanwei@hku.hk).}
\thanks{Z. Ding is with the University of Manchester, Manchester, M1 9BB, UK, and Khalifa University, Abu Dhabi, UAE (e-mail: zhiguo.ding@manchester.ac.uk).}}
\maketitle
\begin{abstract}
Pinching antennas is a novel flexible-antenna technology, which can be realized by employing small dielectric particles on a waveguide. The aim of this letter is to characterize the array gain achieved by pinching-antenna systems (PASS). A closed-form upper bound on the array gain is derived by fixing the inter-antenna spacing. Asymptotic analyses of this bound are conducted by considering an infinitely large number of antennas, demonstrating the existence of an optimal number of antennas that maximizes the array gain. To approach this bound, an antenna position refinement method is introduced. The relationship between the array gain and inter-antenna spacing is further explored by incorporating the effect of mutual coupling. It is proven that there also exists an optimal inter-antenna spacing that maximizes the array gain. Numerical results demonstrate that by optimizing the number of antennas and inter-antenna spacing, PASS can achieve a significantly larger array gain than conventional-antenna systems.
\end{abstract}
\section{Introduction}
Recently, flexible-antenna systems, such as fluid antennas \cite{wong2020fluid} and movable antennas \cite{zhu2024movable}, have gained significant attention. These novel antenna array architecture improve system performance by adjusting the positions of antennas to reconfigure wireless channel conditions. However, conventional flexible-antenna systems face certain limitations. Specifically, antenna movements are often restricted to an aperture on an order of several wavelengths, which makes them ineffective at combating large-scale path loss. Besides, many of these systems are costly to build, and their flexibility in modifying the array structure (e.g., adding or removing antennas) is limited.

To address these challenges, DOCOMO has introduced the \emph{pinching antenna} as a novel flexible-antenna technology \cite{suzuki2022pinching}. Using a dielectric waveguide as the transmission medium, antennas in a \emph{pinching-antenna system (PASS)} can be dynamically activated at any point along the waveguide, much like adding or releasing a clothespin from a clothesline \cite{suzuki2022pinching,ding2024flexible}. This design enables highly flexible and scalable antenna deployment. Unlike conventional flexible-antenna systems, the length of the waveguide can be arbitrarily long, allowing the deployment of pinching antennas very close to the user to establish a strong line-of-sight (LoS) link. Furthermore, a PASS is inexpensive and easy to install, as it only requires adding or removing dielectric materials. In essence, PASS can be viewed as a specific implementation of fluid-antenna or movable-antenna concepts \cite{new2024tutorial}, which offers a more flexible and scalable solution than traditional architectures. In recognition of DOCOMO's original contribution \cite{suzuki2022pinching}, we refer to this technology as \emph{PASS} throughout this paper.

Due to their unique properties, PASS-assisted communications have garnered increasing attention. The pioneering work in \cite{ding2024flexible} analyzed the ergodic rate achieved by employing pinching antennas to serve mobile users and theoretically characterized the performance gain of PASS over conventional fixed-position antenna systems. Following this work, several algorithms have been proposed to optimize the activated positions of pinching antennas along the waveguide \cite{xu2024rate,wang2024antenna,tegos2024minimum}.

Building on existing studies, this article aims to deepen the understanding of PASS by analyzing its achievable array gain and address two fundamental questions about its behavior:
\begin{enumerate}
  \item For fixed inter-antenna spacing, does the array gain \emph{increase monotonically} with the number of antennas?
  \item For a fixed antenna number, does the array gain \emph{increase monotonically} as the inter-antenna spacing decreases?
\end{enumerate}
At first glance, both answers appear to be ``yes'', since it is intuitive to expect that using more antennas would enhance array gain, and reducing inter-antenna spacing would mitigate path loss to the user. However, the analytical findings presented in this work reveal that \emph{the answers to both questions are `no''}.

The main contributions are summarized as follows: \romannumeral1) By fixing the inter-antenna spacing, we derive a closed-form upper bound on the array gain. Using this bound, we prove that the array gain does not always increase monotonically with the number of pinching antennas. Instead, an optimal antenna number exists that maximizes the array gain. We also introduce an antenna position refinement method to approach this bound. \romannumeral2) For a fixed number of antennas, we analyze the impact of inter-antenna spacing on the array gain while considering the effect of mutual coupling (MC) \cite{ivrlavc2010toward}. For the case of two antennas, we derive a closed-form approximation of the array gain. Our analysis demonstrates that, due to MC, reducing the inter-antenna spacing does not guarantee a monotonic array gain improvement. \romannumeral3) We provide numerical results to identify the optimal number of antennas and the optimal inter-antenna spacing. These results highlight the performance superiority of PASS over conventional-antenna systems.
\section{System Model}\label{Section: System Model}
In a downlink communication system, a base station (BS) serves a single-antenna user located at ${\mathbf{u}}=[x_{\rm{u}},0,0]^{\mathsf{T}}$; see {\figurename} {\ref{Figure1}}. Given that PASS is a promising technology for high-frequency bands \cite{suzuki2022pinching}, where LoS propagation typically dominates \cite{ouyang2024primer}, we adopt a free-space LoS channel model to theoretically investigate the performance limits of PASS. The effects of multipath fading will be explored in future work.
\subsection{Conventional-Antenna System}
We first review the conventional \emph{fixed-antenna} system, where the BS antenna is deployed at a fixed location at height $d$, with its position denoted as ${\bm\psi}_{\rm{f}}=[x_{\rm{f}},0,d]^{\mathsf{T}}$. Based on the spherical-wave channel model, the LoS channel coefficient between the fixed antenna and the user is given by \cite{ding2024flexible,xu2024rate,wang2024antenna,tegos2024minimum}
\begin{align}
h_{\rm{f}}=\frac{\eta^{\frac{1}{2}}{\rm{e}}^{-{\rm{j}}k_0\lVert{\mathbf{u}}-{\bm\psi}_{\rm{f}}\rVert}}{\lVert{\mathbf{u}}-{\bm\psi}_{\rm{f}}\rVert}
=\frac{\eta^{\frac{1}{2}}{\rm{e}}^{-{\rm{j}}k_0\sqrt{(x_{\rm{u}}-x_{\rm{f}})^2+d^2}}}{\sqrt{(x_{\rm{u}}-x_{\rm{f}})^2+d^2}},
\end{align}
where $\lVert\cdot\rVert$ denotes the Euclidean norm, $\eta=\frac{c^2}{16\pi^2 f_{\rm{c}}^2}$, $c$ denotes the speed of light, $f_{\rm{c}}$ is the carrier frequency, $k_0=\frac{2\pi}{\lambda}$ is the wavenumber, and $\lambda$ is the free-space wavelength. For the conventional \emph{fluid-antenna} system, the antenna location is denoted as ${\bm\psi}_{\rm{fl}}=[x_{\rm{fl}},0,d]^{\mathsf{T}}$, where the antenna is allowed to be positioned within a spatial region $x_{\rm{fl}}\in[x_{\min},x_{\max}]$ to reconfigure wireless channel conditions \cite{new2024tutorial}. Typically, the range $x_{\max}-x_{\min}$ spans several wavelengths. We have assumed that there is only one single antenna in the conventional-antenna system, not only because it is costly to add more antennas for the conventional case, but also because there is simply no such flexibility to the conventional case \cite{ding2024flexible}. 
\subsection{Pinching-Antenna System}
\begin{figure}[!t]
\centering
\includegraphics[height=0.14\textwidth]{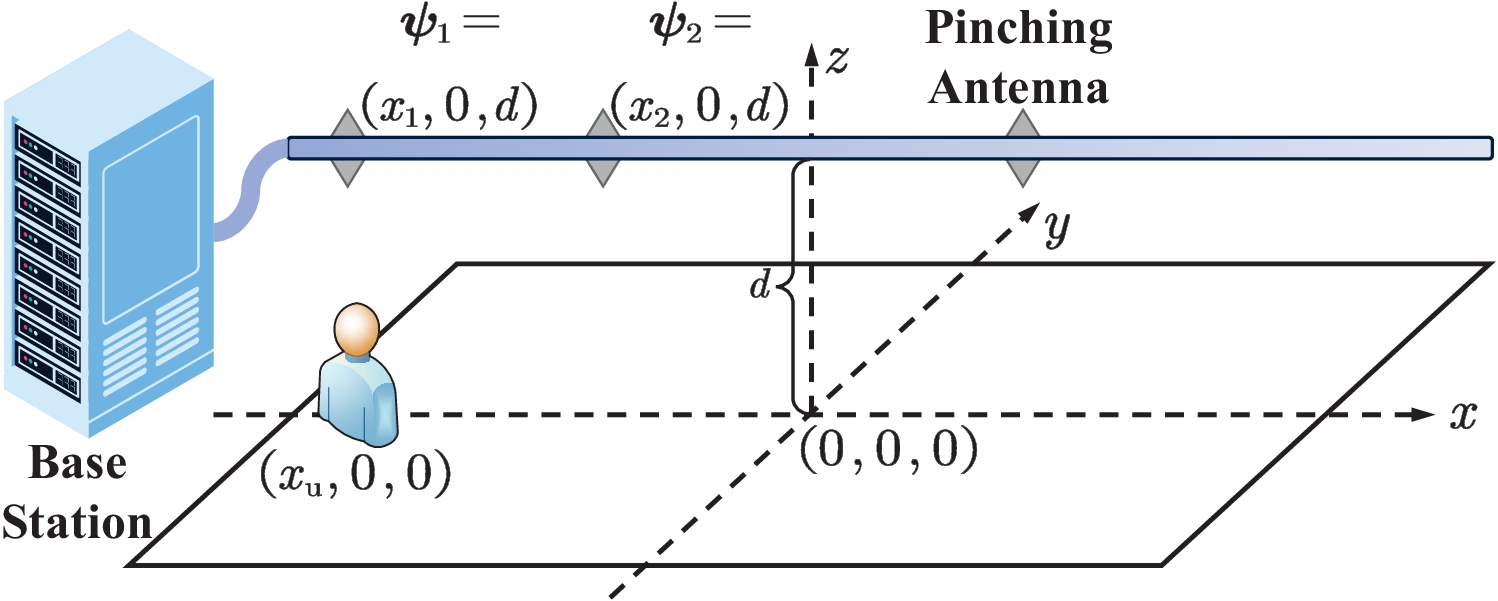}
\caption{Illustration of a PASS with a single waveguide.}
\label{Figure1}
\vspace{-10pt}
\end{figure}

For the PASS, we assume $N$ pinching antennas are activated on a waveguide to jointly serve the user, as depicted in {\figurename} {\ref{Figure1}}. The waveguide is aligned parallel to the $x$-axis at a height $d$. For notational simplicity, $N$ is assumed to be an even integer. The spatial channel coefficient between the $n$th pinching antenna and the user is expressed as follows:
\begin{align}
h_n=\frac{\eta^{\frac{1}{2}}{\rm{e}}^{-{\rm{j}}k_0\lVert{\mathbf{u}}-{\bm\psi}_n\rVert}}{\lVert{\mathbf{u}}-{\bm\psi}_n\rVert}
=\frac{\eta^{\frac{1}{2}}{\rm{e}}^{-{\rm{j}}k_0\sqrt{(x_{\rm{u}}-x_n)^2+d^2}}}{\sqrt{(x_{\rm{u}}-x_n)^2+d^2}},
\end{align}
where ${\bm\psi}_n=[x_n,0,d]^{\mathsf{T}}$ denotes the location of the $n$th pinching antenna for $n\in{\mathcal{N}}\triangleq\{\pm1,\ldots,\pm\frac{N}{2}\}$. Without loss of generality, we consider $x_{n}>x_{n'}$ for $n>n'$. 

Let $s\in{\mathbbmss{C}}$ denote the normalized signal transmitted through the waveguide. The received signal at the user is given by
\begin{align}\label{Signal_Model_Scalar}
y=\sum\nolimits_{n\in{\mathcal{N}}}\sqrt{{P}/{N}}h_n{\rm{e}}^{-{\rm{j}}{\phi_n}}s+z=\sqrt{{P}/{N}}{\mathbf{h}}^{\mathsf{T}}{\bm\phi}s+z,
\end{align}
where ${\bm\phi}=[{\rm{e}}^{-{\rm{j}}\phi_{-\frac{N}{2}}},\ldots,{\rm{e}}^{-{\rm{j}}{\phi_{\frac{N}{2}}}}]^{\mathsf{T}}\in{\mathbbmss{C}}^{N\times1}$, ${\mathbf{h}}=[h_{-\frac{N}{2}},\ldots,h_{\frac{N}{2}}]^{\mathsf{T}}\in{\mathbbmss{C}}^{N\times1}$, $z\sim{\mathcal{CN}}(0,\sigma^2)$ is additive Gaussian noise with variance $\sigma^2$, and $P$ is the total transmit power. The per-antenna power is $\frac{P}{N}$, equally distributed among the $N$ active pinching antennas \cite{ding2024flexible}. The term $\phi_n=\frac{2\pi\lVert{\bm\psi}_n-{\bm\psi}_0\rVert}{\lambda_{\rm{g}}}=\frac{2\pi(x_n-x_0)}{\lambda_{\rm{g}}}$ denotes the in-waveguide phase shift for the $n$th pinching antenna, where ${\bm\psi}_0=[x_0,0,d]^{\mathsf{T}}$ represents the location of the waveguide's feed point with $x_0\leq x_{-\frac{N}{2}}$, $\lambda_{\rm{g}}=\frac{\lambda}{n_{\rm{eff}}}$ is the guided wavelength, and $n_{\rm{eff}}$ is the effective refractive index of the dielectric waveguide \cite{pozar2021microwave}. Propagation loss within the waveguide is neglected in this model due to its limited impact on overall system performance, as justified in prior work \cite{ding2024flexible,wang2024antenna}. This assumption will be further examined in the simulation section. 

Compared with \emph{conventional-antenna} systems, \emph{pinching antennas} can move along the entire length of the waveguide and support scalable deployment by allowing additional antennas to be easily integrated \cite{suzuki2022pinching,ding2024flexible}. Based on \eqref{Signal_Model_Scalar}, the user's signal-to-noise ratio (SNR) for decoding $s$ is given by $\gamma=\frac{P}{\sigma^2}\frac{\lvert{\mathbf{h}}^{\mathsf{T}}{\bm\phi}\rvert^2}{N}$. The array gain achieved by the PASS is given as follows:
\begin{align}\label{PASS_Array_Gain_Basic_Expression}
a=\frac{\gamma}{P/\sigma^2}
=\frac{\eta}{N}\left\lvert\sum_{n\in{\mathcal{N}}}\frac{{{\rm{e}}^{-{\rm{j}}k_0\sqrt{d^2+\Delta_n^2}-{\rm{j}}{k_0\Delta_n}n_{\rm{eff}}}}}{\sqrt{d^2+\Delta_n^2}}\right\rvert^2,
\end{align}
where $\Delta_n\triangleq x_n-x_{\rm{u}}$ for $n\in{\mathcal{N}}$. Since both the received SNR and communication rate are proportional to the array gain, we adopt array gain as the performance metric for the PASS-assisted downlink channel.

\section{Analysis of the Array Gain}
\subsection{Array Gain Versus Antenna Number}\label{Section: Array Gain w.r.t. the Antenna Number}
This section addresses the first question by analyzing how the array gain scales with the number of antennas in a PASS. For brevity, the MC effects between the pinching antennas are neglected. This simplification is achieved by maintaining a minimum inter-antenna spacing of $\Delta_{\rm{p}}\lambda$, where $\Delta_{\rm{p}}\geq \frac{1}{2}$ \cite{ivrlavc2010toward}. 

By \cite[Lemma 2]{xu2024rate}, to maximize the array gain in a single-user PASS, the center of the pinching-antenna array should be positioned directly above the user. This spatial symmetry condition implies $\frac{x_{n}+x_{-n}}{2}=x_{\rm{u}}$ or $\Delta_n=x_n - x_{\rm{u}} =x_{\rm{u}}-x_{-n}=-\Delta_{-n}$ for $n\in{\mathcal{N}}$, which, together with \eqref{PASS_Array_Gain_Basic_Expression}, yields
\begin{align}\label{Array_Gain_Step_Zero}
a=\frac{\eta}{N}\left\lvert\sum_{n=1}^{{N}/{2}}\frac{2{{\rm{e}}^{-{\rm{j}}k_0d\sqrt{1+\Delta_n^2/d^2}}}\cos({k_0\Delta_n}n_{\rm{eff}})}{d\sqrt{1+\Delta_n^2/d^2}}\right\rvert^2
\triangleq a_{\rm{nui}}.
\end{align}
\subsubsection{Array Gain with Equal Spacing}
For the case where the $N$ antennas are equally spaced, we have $\Delta_n= (n-\frac{1}{2}) \Delta_{\rm{p}}\lambda $ for $n\geq1$. Substituting this expression into \eqref{Array_Gain_Step_Zero} yields
\begin{align}\label{Array_Gain_Step_First}
a_{\rm{nui}}=\frac{\eta}{N}\left\lvert\sum\nolimits_{n=1}^{{N}/{2}}\frac{1}{\lambda}{f_{\rm{a}}((n  - 1/2)\varepsilon)\varepsilon}\right\rvert^2\triangleq a_{\rm{uni}},
\end{align} 
where $\varepsilon\triangleq\frac{\lambda}{d}$ and $f_{\rm{a}}(x)\triangleq\frac{2{{\rm{e}}^{-{\rm{j}}k_0d\sqrt{1+\Delta_{\rm{p}}^2x^2}}}\cos({k_0d\Delta_{\rm{p}}}n_{\rm{eff}}x)}{\sqrt{1+\Delta_{\rm{p}}^2x^2}}$. Since $\varepsilon\ll1$, the summation in \eqref{Array_Gain_Step_First} can be accurately approximated using a definite integral as follows \cite{ouyang2024primer}:
\begin{align}\label{Array_Gain_Fixed_Spacing_General1}
a_{\rm{uni}}\approx\frac{\eta\big\lvert \int_{0}^{{N\varepsilon}/{2}}f_{\rm{a}}(x){\rm{d}}x\big\rvert^2}{N\lambda^2}=\frac{\eta\big\lvert \int_{0}^{{N\varepsilon}/{2}}f_{\rm{a}}(x){\rm{d}}x\big\rvert^2}{Nd^2\varepsilon^2}.
\end{align} 
\subsubsection{An Upper Bound of the Array Gain}
The complexity introduced by the term ${{\rm{e}}^{-{\rm{j}}k_0d\sqrt{1+\Delta_n^2/d^2}}}\cos({k_0\Delta_n}n_{\rm{eff}})$ makes it challenging to extract further insights from \eqref{Array_Gain_Step_Zero} or \eqref{Array_Gain_Step_First}. As a compromise, we construct an upper bound on \eqref{Array_Gain_Step_Zero} as follows:
\begin{align}\label{General_Upper_Bound_Array_Gain}
a_{\rm{nui}}\leq\frac{\eta}{N}\bigg\lvert\sum\nolimits_{n=1}^{{N}/{2}}\frac{2}{d\sqrt{1+\Delta_n^2/d^2}}\bigg\rvert^2\triangleq \hat{a}_{\rm{nui}}.
\end{align}
Inserting $\Delta_n= (n-\frac{1}{2}) \Delta_{\rm{p}}\lambda $ ($n\geq1$) into the right-hand side (RHS) of \eqref{General_Upper_Bound_Array_Gain} gives the upper bound on \eqref{Array_Gain_Step_First} as follows:
\begin{align}
a_{\rm{uni}}\leq
\frac{\eta}{N}\Bigg\lvert\sum_{n=1}^{{N}/{2}}\frac{2}{d\sqrt{1+{((n-\frac{1}{2}) \Delta_{\rm{p}}\lambda)^2}/d^2}}\Bigg\rvert^2\triangleq \hat{a}_{\rm{uni}}.\label{Array_Gain_Fixed_Spacing_General2b}
\end{align}
Similar to the derivation of \eqref{Array_Gain_Fixed_Spacing_General1}, we obtain
\begin{subequations}
\begin{align}
\hat{a}_{\rm{uni}}&\approx\frac{\eta}{Nd^2\varepsilon^2}\bigg\lvert \int_{0}^{\frac{N\varepsilon}{2}}\frac{2}{\sqrt{1+\Delta_{\rm{p}}^2x^2}}{\rm{d}}x\bigg\rvert^2\\
&=\frac{4\eta(\ln(\sqrt{1+L_{\varepsilon}^2}+L_{\varepsilon}))^2}{\Delta_{\rm{p}}^2Nd^2\varepsilon^2}=\frac{2\eta f_{\rm{ub}}\left(L_{\varepsilon}\right)}{\Delta_{\rm{p}}d^2\varepsilon},\label{Upper_Bound_Calculation_Integral}
\end{align}
\end{subequations}
where $L_{\varepsilon}\triangleq\frac{N\Delta_{\rm{p}}\varepsilon}{2}$ and $f_{\rm{ub}}(x)\triangleq\frac{(\ln(\sqrt{1+x^2}+x))^2}{x}$. 
\begin{theorem}\label{Theorem1}
Given $\Delta_{\rm{p}}\geq\frac{1}{2}$, as $N\rightarrow\infty$, the array gain of the $N$ equally spaced antennas satisfies $\lim_{N\rightarrow\infty}a_{\rm{uni}}=0$.
\end{theorem}
\begin{IEEEproof}
From \eqref{Array_Gain_Fixed_Spacing_General2b}, $0\leq a_{\rm{uni}}\leq \frac{2\eta f_{\rm{ub}}\left(L_{\varepsilon}\right)}{\Delta_{\rm{p}}d^2\varepsilon}$. Since $\lim_{x\rightarrow\infty}f_{\rm{ub}}(x)=0$, it has $\lim_{N\rightarrow\infty} \frac{2\eta f_{\rm{ub}}\left(L_{\varepsilon}\right)}{\Delta_{\rm{p}}d^2\varepsilon}=0$. Applying the squeeze theorem, we conclude that $\lim_{N\rightarrow\infty}a_{\rm{uni}}=0$.
\end{IEEEproof}
The results in {Theorem \ref{Theorem1}} can be interpreted as follows. As $N$ tends to infinity, the power per antenna, i.e., $\frac{P}{N}$, decreases. Therefore, the antennas that account for the majority of the power are positioned too far from the user, which makes the user receive negligible energy from the pinching antennas.
\begin{remark}\label{remark_number}
This finding implies that increasing the number of pinching antennas does not guarantee a continuous increase in array gain for the considered PASS. Instead, there exists an optimal number of antennas that maximizes the array gain.
\end{remark}
Since the term $\frac{1}{\sqrt{\Delta_n^2+d^2}}$ declines as $\Delta_n$ increases, the RHS of \eqref{Array_Gain_Fixed_Spacing_General2b}, i.e., $\hat{a}_{\rm{uni}}$, also serves as an upper bound for $a_{\rm{nui}}$ in \eqref{Array_Gain_Step_Zero}, i.e., $\hat{a}_{\rm{uni}}\leq a_{\rm{nui}}$, provided that there is a minimum inter-antenna spacing of $\Delta_{\rm{p}}\lambda$ (i.e., $\lvert\Delta_{n}-\Delta_{n'}\rvert\geq\Delta_{\rm{p}}\lambda$ for $n\ne n'$) and MC effects are neglected (i.e., $\Delta_{\rm{p}}\geq\frac{1}{2}$). Therefore, the conclusion in Remark \ref{remark_number} is also applicable to a non-uniformly spaced PASS, as long as MC effects are ignored and the spacing condition is met.
\subsubsection{A Method to Approach the Upper Bound}\label{Section: A Method to Approach the Upper Bound}
After establishing the upper bound of the array gain, we apply the methodology from \cite{xu2024rate} to approach it. The upper bound in \eqref{Array_Gain_Fixed_Spacing_General2b} is derived by minimizing the inter-antenna spacing and neglecting the dual phase shifts induced by signal propagation inside and outside the waveguide, i.e., $k_0(\sqrt{d^2+\Delta_n^2}+{\Delta_n}n_{\rm{eff}})$. To achieve this bound, we must ensure constructive combination of the received signals from distinct pinching antennas at the user. We satisfy this condition by slightly adjusting the location of each pinching antenna, as detailed below. Due to the symmetry between $x_n$ and $x_{-n}$ with respect to (w.r.t.) $x_{\rm{u}}$, we analyze only the case of $n\geq1$ for conciseness.

For the $n$th pinching antenna, we position it to the right of $x_n=x_{\rm{u}}+\Delta_{n}$ by a distance $v_n>0$ to satisfy
\begin{align}\label{Refininement_Equation}
\sqrt{d^2+(\Delta_n+v_n)^2}+{(\Delta_n+v_n)}n_{\rm{eff}}=d_n,
\end{align}
where $\Delta_1=\frac{\Delta_{\rm{p}}\lambda}{2}$, $d_n=\lambda\lceil\frac{1}{\lambda}(\sqrt{d^2+\Delta_n^2}+{\Delta_n}n_{\rm{eff}})\rceil$, and $\lceil\cdot\rceil$ represents the ceiling operator. The solution is given by
\begin{align}\label{Refininement_Solution}
v_n^{\star}=\left\{\begin{matrix}\frac{d_nn_{\rm{eff}}-\sqrt{d_n^2+d^2(n_{\rm{eff}}^2-1)}}{n_{\rm{eff}}^2-1}-\Delta_n&n_{\rm{eff}}\ne 1\\
\frac{d_n^2-d^2}{2d_n}-\Delta_n&n_{\rm{eff}}=1\end{matrix}\right..
\end{align}
Since a propagation distance of one wavelength introduces a $2\pi$-phase shift and the left-hand side of \eqref{Refininement_Equation} increases monotonically with $v_n$, the optimal shift $v_n^{\star}$ is on the wavelength scale. This distance is much smaller than the height $d$, which ensures negligible impact on large-scale path loss.

Once we obtain $v_n^{\star}$, we update $\Delta_n\leftarrow\Delta_n+v_n^{\star}$ and set $\Delta_{n+1}=\Delta_n+\Delta_{\rm{p}}\lambda$ to maintain an inter-antenna spacing of at least $\Delta_{\rm{p}}\lambda$. We substitute $\Delta_{n}=\Delta_{n+1}$ into \eqref{Refininement_Equation} and \eqref{Refininement_Solution} and transform $v_{n}\rightarrow v_{n+1}$ to solve for the refined distance $v_{n+1}^{\star}$ for the $(n+1)$th pinching antenna. We position the remaining pinching antennas sequentially using the same procedure, with their \emph{refined locations} expressed as follows:
\begin{align}\label{Refined_Antenna_Location}
x_{n}=x_{\rm{u}}+\left(n-1/2\right)\Delta_{\rm{p}}\lambda+\sum\nolimits_{i=1}^{n}v_{i}^{\star},~n\geq1.
\end{align}
Equation \eqref{Refined_Antenna_Location} shows that the proposed method closely tracks the upper bound in \eqref{Array_Gain_Fixed_Spacing_General2b} when $d\gg\sum\nolimits_{i=1}^{n}v_{i}^{\star}$, which is a generally mild condition since $v_n^{\star}$ is on the wavelength scale and $d\gg \lambda$. Section \ref{Section_Numerical_Results} numerically validates that this refinement method approximates the upper bound with high accuracy. Further validations and details are available in \cite{xu2024rate}.
\subsubsection{Optimal Antenna Number and Array Gain Limits}\label{Section: Optimal Antenna Number and Array Gain Limits}
Since $\hat{a}_{\rm{uni}}$, the upper bound of \eqref{Array_Gain_Step_Zero}, can be closely tracked using the method in \cite{xu2024rate}, we treat $\hat{a}_{\rm{uni}}$ as an accurate approximation of the maximum achievable array gain, i.e., $\hat{a}_{\rm{uni}}\approx \max_{\lvert\Delta_{n}-\Delta_{n'}\rvert\geq\Delta_{\rm{p}}\lambda,n\ne n'} a_{\rm{nui}}$, to gain more insights.

By examining the derivative of $f_{\rm{ub}}(x)$ w.r.t. $x$ and numerically solving $\frac{{\rm{d}}}{{\rm{d}}x}f_{\rm{ub}}(x)=0$ using the bisection method, we find that $f_{\rm{ub}}(x)$ is maximized at $x=x^{\star}\approx3.32$, with $f_{\rm{ub}}(x^{\star})\approx1.105$, as shown in {\figurename} {\ref{fig3a}}. This implies that the optimal number of pinching antennas $N^{\star}$ satisfies
\begin{align}\label{Exp_Optimal_Antenna_Number}
\frac{N^{\star}\Delta_{\rm{p}}\varepsilon}{2}\approx x^{\star}\Leftrightarrow N^{\star}\approx\frac{2x^{\star}}{\Delta_{\rm{p}}\varepsilon}=\frac{2x^{\star}d}{\Delta_{\rm{p}}\lambda}\approx
\frac{6.64d}{\Delta_{\rm{p}}\lambda}.
\end{align} 
While $N^{\star}$ depends on $\{d,\Delta_{\rm{p}},\lambda\}$, the array aperture covered by the $N^{\star}$ pinching antennas is given by
\begin{align}
(N^{\star}-1)\Delta_{\rm{p}}\lambda\approx2x^{\star}d-\Delta_{\rm{p}}\lambda\approx 2x^{\star}d\approx 6.64 d,
\end{align}
which depends primarily on $d$. For example, when $d=1$ m, the aperture is approximately $6.64$ m, and for $d=3$ m, it increases to $19.92$ m. These dimensions are practical for waveguide deployments in indoor environments such as libraries or shopping malls, which underscores the need to optimize the number of pinching antennas. 

We further observe that $f_{\rm{ub}}(x)$ increases monotonically for $x\in(0,x^{\star})$. This implies that leveraging more pinching antennas is advantageous when $(N-1)\Delta_{\rm{p}}\lambda\leq 2x^{\star}d$ or $N\leq N^{\star}$, which enhances the array gain. Setting $N=N^{\star}$ gives the maximum of the upper bound as follows:
\begin{align}\label{Array_Gain_Maximum}
\frac{2\eta }{\Delta_{\rm{p}}d^2\varepsilon}f_{\rm{ub}}\left({N^{\star}\Delta_{\rm{p}}\varepsilon}/{2}\right)\approx\frac{2\eta f_{\rm{ub}}(x^{\star}) }{d\Delta_{\rm{p}}\lambda}
\approx \frac{2.21\eta}{d\Delta_{\rm{p}}\lambda}.
\end{align}
\begin{remark}\label{Remark_Conclusion_Decrease}
Equations \eqref{Exp_Optimal_Antenna_Number} and \eqref{Array_Gain_Maximum} indicate that the optimal antenna number $N^{\star}$ and the maximum array gain decrease as the minimum inter-antenna spacing $\Delta_{\rm{p}}$ increases. 
\end{remark}
\begin{remark}
Since \eqref{Array_Gain_Maximum} is derived for $\Delta_{\rm{p}}\geq\frac{1}{2}$ (to suppress MC), we conclude that the maximum achievable array gain cannot exceed $\frac{2\eta f_{\rm{ub}}(x^{\star}) }{d(\lambda/2)}\approx \frac{4.42\eta}{d\lambda}$, i.e., $\frac{2\eta f_{\rm{ub}}\left({N\Delta_{\rm{p}}\varepsilon}/{2}\right)}{\Delta_{\rm{p}}d^2\varepsilon}\leq\frac{4.42\eta}{d\lambda}$. This yields the upper limit of the array gain achieved by PASS.
\end{remark}
\subsection{Array Gain Versus Antenna Spacing}\label{Section: Array Gain w.r.t. the Antenna Spacing}
After addressing the first question, we investigate the second question concerning the relationship between array gain and inter-antenna spacing. Equation \eqref{General_Upper_Bound_Array_Gain} indicates that reducing the inter-antenna spacing $\lvert \Delta_n-\Delta_{n'}\rvert$ ($n\ne n'$) enhances the array gain. Equation \eqref{Array_Gain_Maximum} further suggests that setting $\Delta_{\rm{p}}\lambda=0$ or $\Delta_n=0$ could theoretically yield an infinitely large array gain. However, this result violates energy conservation and is therefore unphysical. The discrepancy arises because \eqref{General_Upper_Bound_Array_Gain} and \eqref{Array_Gain_Maximum} assume $\Delta_{\rm{p}}\geq \frac{1}{2}$ and neglect MC effects. When $\lvert \Delta_n-\Delta_{n'}\rvert$ becomes arbitrarily small, MC effects dominate \cite{ivrlavc2010toward}, and \eqref{General_Upper_Bound_Array_Gain} and \eqref{Array_Gain_Maximum} fail to accurately predict PASS performance. 

MC occurs when electromagnetic waves transmitted by one antenna are absorbed by adjacent antennas, perturbing their circuitry and distorting the spatial channel. To rigorously evaluate the array gain versus inter-antenna spacing, we incorporate MC effects. The spatial channel between the $N$ pinching antennas and the user becomes ${\mathbf{g}}={\mathbf{C}}^{-\frac{1}{2}}{\mathbf{h}}={\mathbf{C}}^{-\frac{1}{2}}[h_{-{N}/{2}},\ldots,h_{{{N}/{2}}}]^{\mathsf{T}}$ \cite[Eq. (105)]{ivrlavc2010toward}, where ${\mathbf{C}}\in{\mathbbmss{C}}^{N\times N}$ denotes the MC matrix. For simplicity, we assume that the pinching antennas are evenly spaced with spacing $\Delta$ and $\frac{x_{n}+x_{-n}}{2}=x_{\rm{u}}$. Based on \cite[Eq. (48)]{ivrlavc2010toward}, we model ${\mathbf{C}}$ as ${\mathbf{C}}=\begin{bmatrix}\begin{smallmatrix}
               1 & J(2) & \ldots & J(N)\\
               J(2) & 1 & \ldots & J(N-1)\\
               \vdots & \ddots & \ddots & \vdots\\
               J(N) & J(N-1) & \ldots & 1
             \end{smallmatrix}\end{bmatrix}$, where $J(n)\triangleq\frac{\sin(k_0\Delta(n-1))}{k_0\Delta(n-1)}$ for $n=1,\ldots,N$. When $\Delta=\frac{\lambda}{2}$, it holds that ${\mathbf{C}}={\mathbf{I}}_N$, which simplifies ${\mathbf{g}}={\mathbf{h}}$ and eliminates MC effects \cite{ivrlavc2010toward}.

\begin{figure}[!t]
    \centering
    \subfloat[$f_{\rm{ub}}(x)$.]
    {
        \includegraphics[height=0.172\textwidth]{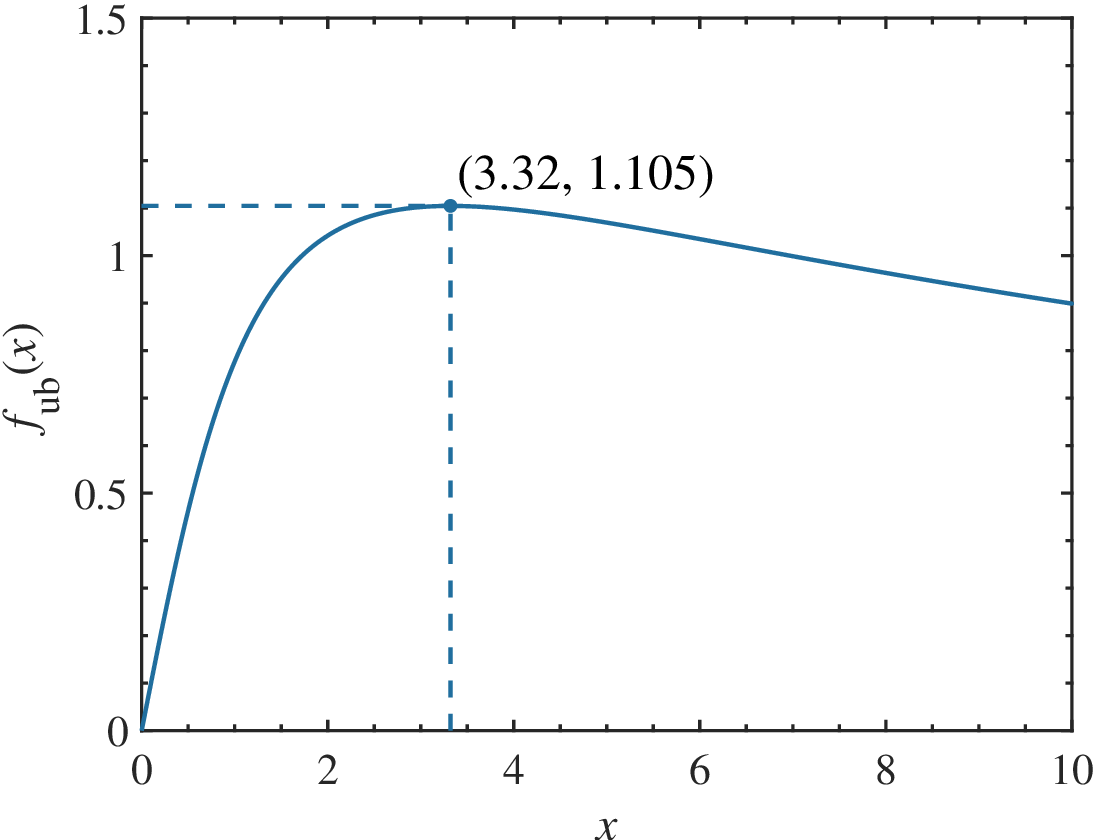}
	   \label{fig3a}	
    }
   \subfloat[$f_{\rm{mc}}({\Delta}/{\lambda})$.]
    {
        \includegraphics[height=0.172\textwidth]{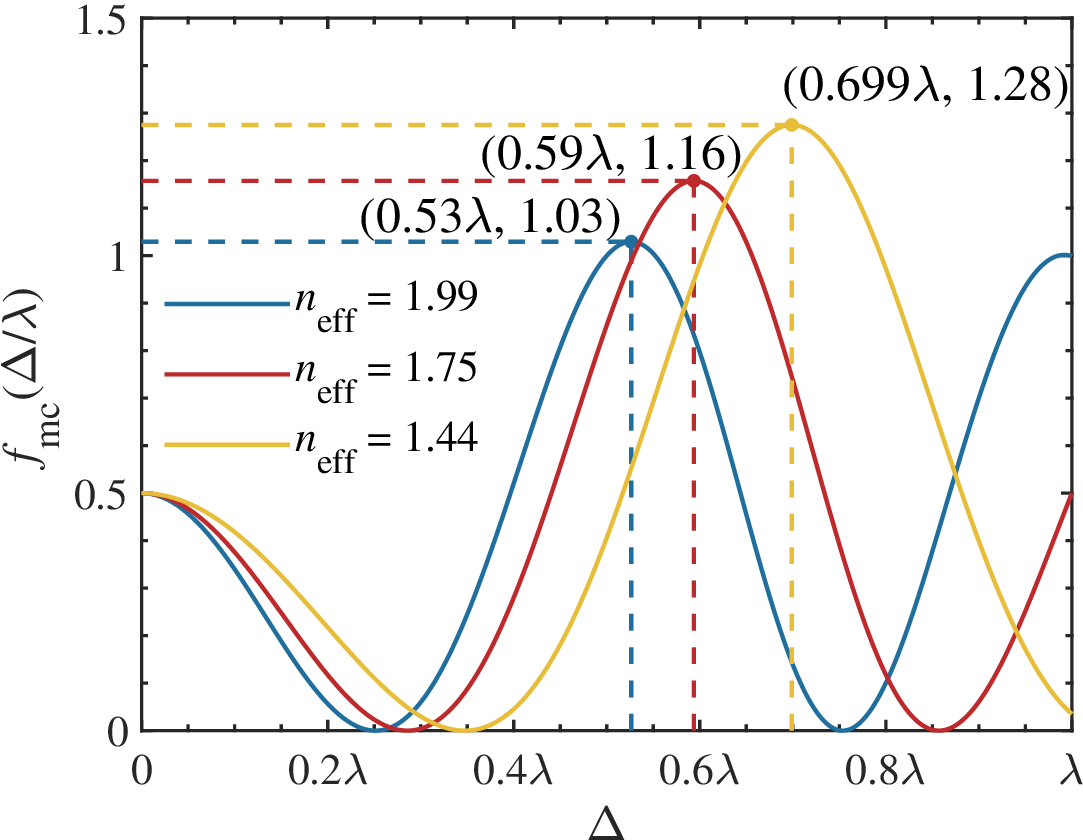}
	   \label{fig3b}	
    }
\caption{Illustration of $f_{\rm{ub}}(x)$ and $f_{\rm{mc}}({\Delta}/{\lambda})$.}
    \label{Figure3}
    \vspace{-10pt}
\end{figure}

\begin{figure*}[!t]
    \centering
    \subfloat[Array gain without MC.]
    {
        \includegraphics[height=0.22\textwidth]{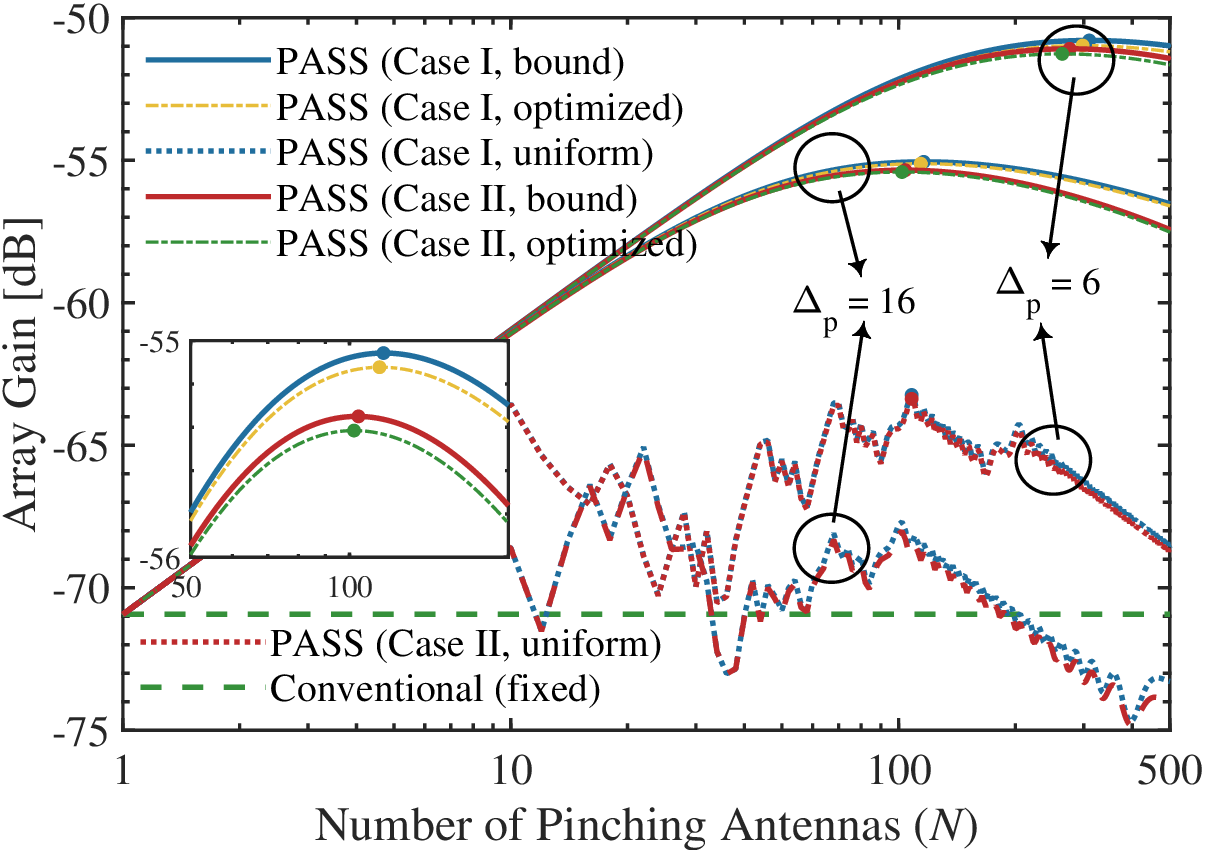}
	   \label{fig4a}	
    }
   \subfloat[Maximum array gain without MC.]
    {
        \includegraphics[height=0.22\textwidth]{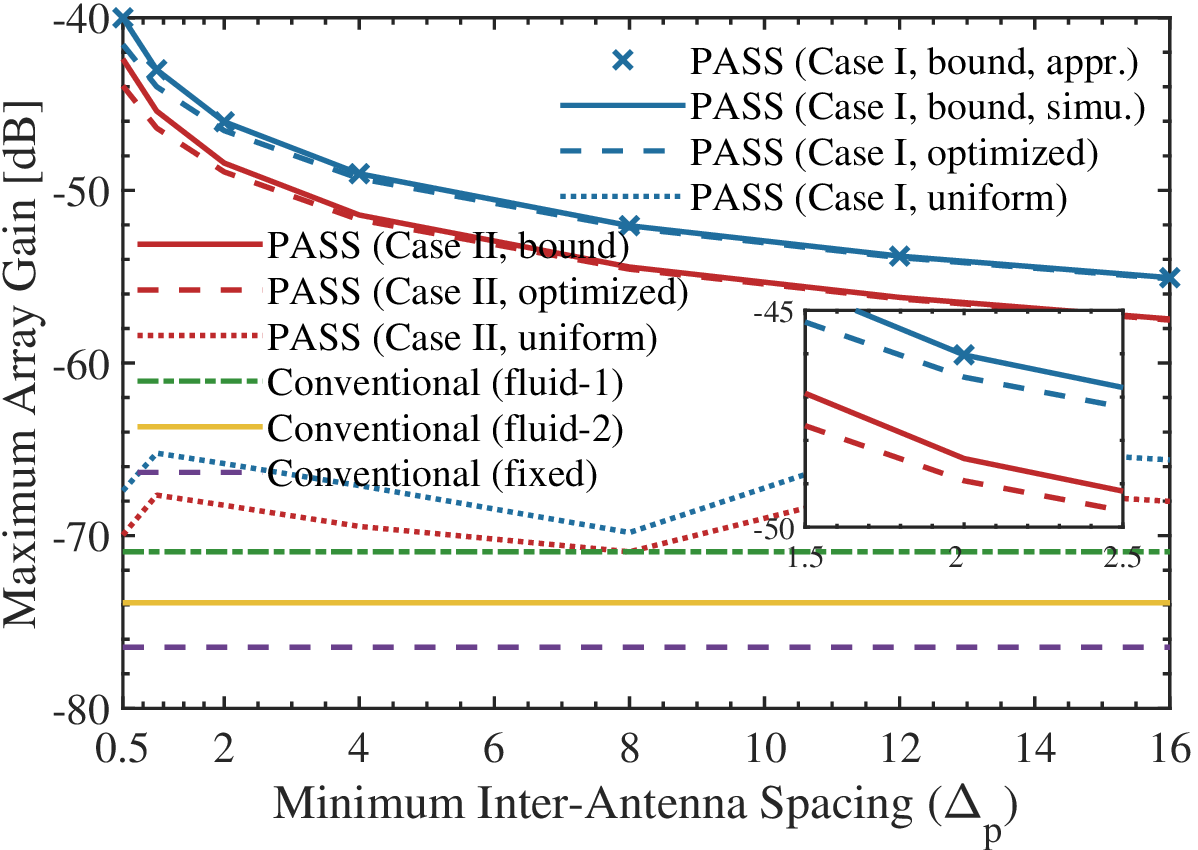}
	   \label{fig4b}	
    }
    \subfloat[Array gain with/without MC.]
    {
        \includegraphics[height=0.22\textwidth]{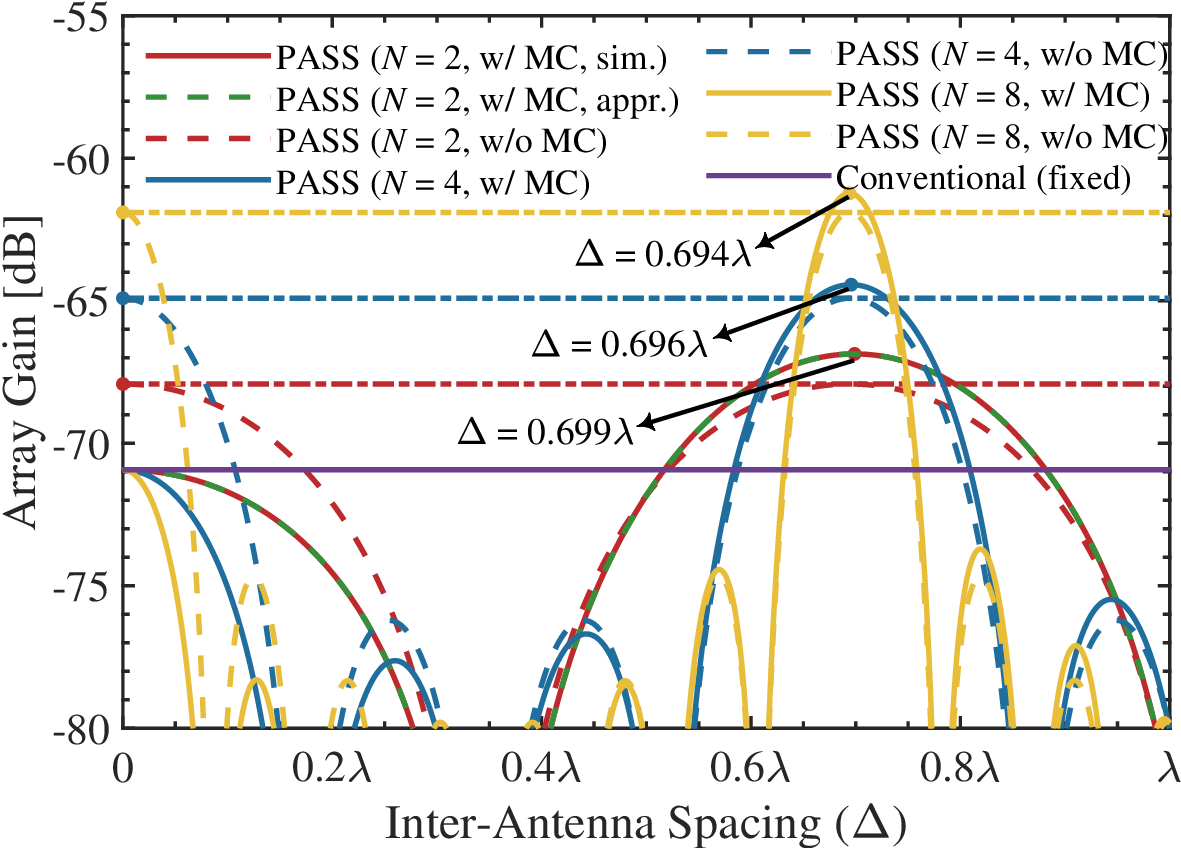}
	   \label{fig4c}	
    }
\caption{The array gain achieved by the pinching antennas, where the maximum array gain ($\bullet$) in {\figurename} {\ref{fig4a}} and {\figurename} {\ref{fig4c}} is obtained via a one-dimensional search.}
    \label{Figure4}
    \vspace{-15pt}
\end{figure*}

As a result, the array gain can be written as follows:
\begin{align}\label{MC_Array_Gain_Basic_Exp}
a={\lvert{\mathbf{g}}^{\mathsf{T}}{\bm\phi}\rvert^2}/{N}={\lvert{\mathbf{h}}^{\mathsf{T}}{\mathbf{C}}^{-\frac{1}{2}}{\bm\phi}\rvert^2}/{N}\triangleq a_{\rm{mc}}.
\end{align}
Direct computation of ${\mathbf{C}}^{-\frac{1}{2}}$ is generally intractable. We thus analyze the simplified case of two closely spaced antennas ($N=2$) at positions $x_1=x_{\rm{u}}+\frac{\Delta}{2}$ and $x_{-1}=x_{\rm{u}}-\frac{\Delta}{2}$, where the MC matrix reduces to $\begin{bmatrix}\begin{smallmatrix}
               1 & J(2)\\
               J(2) & 1
             \end{smallmatrix}\end{bmatrix}\triangleq{\mathbf{C}}_2$. By \cite[Eqs. (1)–(3)]{deledalle2017closed}, the eigendecomposition of ${\mathbf{C}}_2$ is given by
\begin{align}
{\mathbf{C}}_2=\left[\begin{matrix}\frac{-1}{\sqrt{2}}&\frac{-1}{\sqrt{2}}\\\frac{-1}{\sqrt{2}}&\frac{1}{\sqrt{2}}\end{matrix}\right]
\left[\begin{matrix}1+J(2)&0\\0&1-J(2)\end{matrix}\right]
\left[\begin{matrix}\frac{-1}{\sqrt{2}}&\frac{-1}{\sqrt{2}}\\\frac{-1}{\sqrt{2}}&\frac{1}{\sqrt{2}}\end{matrix}\right].
\end{align}
Consequently, ${\mathbf{C}}_2^{-\frac{1}{2}}=\left[\begin{smallmatrix}\frac{-1}{\sqrt{2}}&\frac{-1}{\sqrt{2}}\\\frac{-1}{\sqrt{2}}&\frac{1}{\sqrt{2}}\end{smallmatrix}\right]
\left[\begin{smallmatrix}\frac{1}{\sqrt{1+J(2)}}&0\\0&\frac{1}{\sqrt{1-J(2)}}\end{smallmatrix}\right]
\left[\begin{smallmatrix}\frac{-1}{\sqrt{2}}&\frac{-1}{\sqrt{2}}\\\frac{-1}{\sqrt{2}}&\frac{1}{\sqrt{2}}\end{smallmatrix}\right]$. For $N=2$ with $\Delta_1=\Delta_{-1}=\frac{\Delta}{2}$, we have $h_1=h_{-1}=\frac{\eta^{1/2}{\rm{e}}^{-{\rm{j}}\sqrt{d^2+{\Delta^2}/{4}}}}{\sqrt{d^2+{\Delta^2}/{4}}}$ and $\phi_1=-\phi_{-1}={\pi n_{\rm{eff}}}{\frac{\Delta}{\lambda}}$. Substituting these into \eqref{MC_Array_Gain_Basic_Exp}, the array gain becomes
\begin{align}
a_{\rm{mc}}&=\frac{{\lvert{\mathbf{h}}^{\mathsf{T}}{\mathbf{C}}_2^{-{1}/{2}}{\bm\phi}\rvert^2}}{2}
=\frac{1}{8}\bigg\lvert\frac{(h_1+h_{-1})({\rm{e}}^{-\rm{j}\phi_{-1}}+{\rm{e}}^{-\rm{j}\phi_{1}})}{\sqrt{1+j_0(k_0\Delta)}}\bigg\rvert^2\nonumber\\
&=\frac{2\eta\cos^2(\frac{n_{\rm{eff}}}{2}{k_0\Delta})}{({d^2+\frac{\Delta^2}{4}})({1+j_0(k_0\Delta)})}\approx
\frac{2\eta\cos^2(\frac{n_{\rm{eff}}}{2}{k_0\Delta})}{d^2({1+j_0(k_0\Delta)})}\label{Array_Gain_Distance_2nd},
\end{align} 
where ${\Delta^2}/{4}$ in the denominator is neglected due to $\Delta\ll d$.
\begin{remark}
At $\Delta=0$, \eqref{Array_Gain_Distance_2nd} simplifies to $a_{\rm{mc}}=\frac{\eta}{d^2}$. By contrast, neglecting MC in \eqref{PASS_Array_Gain_Basic_Expression} gives $a=\frac{2\eta}{d^2}$. This demonstrates that MC reduces the array gain, which underscores the necessity of modeling MC effects in array gain analysis.
\end{remark}
To gain further insights, we substitute $k_0=\frac{2\pi}{\lambda}$ into \eqref{Array_Gain_Distance_2nd} and rewrite it as $a_{\rm{mc}}\approx\frac{2\eta}{d^2}f_{\rm{mc}}(\frac{\Delta}{\lambda})$, where $f_{\rm{mc}}(x)\triangleq \frac{\cos^2({\pi n_{\rm{eff}}}x)}{({1+j_0(2\pi x)})}$. {\figurename} {\ref{fig3b}} plots $f_{\rm{mc}}(\frac{\Delta}{\lambda})$ w.r.t. $\Delta\in[0,\lambda]$ for $n_{\rm{eff}}$ values from \cite{rodrigues2023all}. The non-monotonic behavior of $f_{\rm{mc}}(\frac{\Delta}{\lambda})$ indicates that reducing $\Delta$ does not always improve the array gain. A one-dimensional search over $\Delta\in[0,\lambda]$ reveals that the maximum $\frac{2\eta}{d^2}f_{\rm{mc}}\left(\frac{\Delta}{\lambda}\right)\approx a_{\rm{mc}}$ exceeds both the MC-neglected array gain $\frac{2\eta}{d^2}$ at $\Delta=0$ and the MC-aware array gains at $\Delta=0$ (i.e., $\frac{\eta}{d^2}$) and $\Delta=\frac{\lambda}{2}$; see {\figurename} {\ref{fig3b}}. This improvement arises from the combined effects of in-waveguide transmission and MC.
\begin{remark}\label{remark_spacing}
In PASS, reducing inter-antenna spacing does not ensure a monotonic increase in array gain. Instead, an optimal inter-antenna spacing exists to maximize the array gain.
\end{remark}
These findings answer the second question and demonstrate that MC in PASS can enhance the array gain. However, the MC matrix typically exhibits complex structure and strongly depends on the pinching antennas' positions. Our analysis is restricted to the two-antenna case due to these complexities, while generalizing to arbitrary numbers of pinching antennas remains an open research challenge. We anticipate that these results will motivate future MC-aware PASS designs.
\section{Numerical Results}\label{Section_Numerical_Results}
This section evaluates the impact of the number of pinching antennas and inter-antenna spacing on array gain via numerical simulations. We use the following parameters unless stated otherwise: $f_{\rm{c}}=28$ GHz, $d=3$ m \cite{ding2024flexible}, $x_{\rm{u}}=0$ m, $x_{\rm{f}}=0$ m, $x_{0}=x_{-\frac{N}{2}}$, and $n_{\rm{eff}}=1.44$ \cite{rodrigues2023all}. We analyze two PASS configurations: \romannumeral1) Case {\uppercase\expandafter{\romannumeral1}}: no waveguide propagation loss and \romannumeral2) Case {\uppercase\expandafter{\romannumeral2}}: waveguide propagation loss of $0.08$ dB/m \cite{ding2024flexible,wang2024antenna}.

{\figurename} {\ref{fig4a}} plots the array gain versus the number of pinching antennas for different $\Delta_{\rm{p}}$ values. The upper bound is computed via \eqref{Array_Gain_Fixed_Spacing_General2b}, while the optimized array gain follows the methodology from Section \ref{Section: A Method to Approach the Upper Bound}. Although \eqref{Upper_Bound_Calculation_Integral} provides a precise approximation of the upper bound, we choose not to display it for simplicity, as its accuracy is well-documented \cite{ouyang2024primer}. For comparison, we also present the channel gains from a uniformly or equally spaced PASS and a conventional fixed-location antenna, which are lower than those achieved by the optimized PASS. Notably, both the array gain achieved by the PASS and its upper bound exhibit non-monotonic dependence on $N$, with an optimal antenna number maximizing the array gain, which corroborates the conclusion in Remark \ref{remark_number}. Additionally, {\figurename} {\ref{fig4a}} indicates that the optimal antenna number and maximum array gain decrease as $\Delta_{\rm{p}}$ increases, which confirms Remark \ref{Remark_Conclusion_Decrease}. The optimized array gain closely tracks the upper bound, validating $\hat{a}_{\rm{uni}}\approx \max_{\lvert\Delta_{n}-\Delta_{n'}\rvert\geq\Delta_{\rm{p}}\lambda,n\ne n'} a_{\rm{nui}}$ and supporting the analysis in Section \ref{Section: Optimal Antenna Number and Array Gain Limits}. The figure also reflects that the effect of waveguide propagation loss is insignificant, which is consistent with \cite{ding2024flexible,wang2024antenna}.

{\figurename} {\ref{fig4b}} plots the maximum array gain versus the minimum inter-antenna spacing, which is obtained via a one-dimensional search over $N\in[0,10^4]$. We assume the user's location $x_{\rm{u}}$ follows a uniform distribution within $[-15~{\text{m}},15~{\text{m}}]$ and average results over $10^3$ channel realizations for $x_0=-30$ m. For comparison, we include the array gain of conventional fluid-antenna systems: \romannumeral1) fluid-1: the fluid antenna aligns directly above the user with $x_{\rm{fl}}=x_{\rm{u}}$ and \romannumeral2) fluid-2: the fluid antenna moves within $x_{\rm{fl}}\in[-500\lambda,500\lambda]$. As illustrated in {\figurename} {\ref{fig4b}}, the approximated bound (computed via \eqref{Array_Gain_Maximum}) closely matches the simulated results. Besides, we observe that in both configurations (with/without waveguide propagation loss), the maximum array gain of PASS decreases with increasing minimum inter-antenna spacing, which confirms Remark \ref{Remark_Conclusion_Decrease}. Notably, both the optimized PASS and the uniformly spaced PASS outperform, or at least match, the array gain of conventional-antenna systems (fluid and fixed). This superiority stems from the high flexibility of PASS in reducing path loss and its scalability in adding more antennas. Moreover, waveguide propagation loss causes only a small degradation in performance. Even with this loss, PASS still achieves significant performance improvements over conventional systems.

{\figurename} {\ref{fig4c}} plots the array gain as a function of inter-antenna spacing $\Delta$ when MC is considered. Given the limited impact of waveguide propagation loss and that $\Delta\in[0,\lambda]$, we present results only for Case {\uppercase\expandafter{\romannumeral1}}. It can be seen that when using two antennas, the simulated array gain closely aligns with the analytical approximation in \eqref{Array_Gain_Distance_2nd}. The graph shows that, in the absence of MC, the array gain reaches its maximum at $\Delta=0$ for all considered $N$ values. This indicates that minimizing inter-antenna spacing can maximize array gain, which is equivalent to enlarging a single antenna's effective aperture. However, when MC is considered, the array gain oscillates with $\Delta$, and an optimal spacing exists to maximize the gain, which is consistent with Remark \ref{remark_spacing}. Besides, the optimal $\Delta$ for $N=2$ in {\figurename} {\ref{fig4c}} coincides with the result shown in {\figurename} {\ref{fig3b}}. Finally, this graph confirms that the considered PASS outperforms conventional fixed-location antennas and that MC can be used to enhance array gain for all considered $N$ values.
\section{Conclusion} 
This article has analyzed the array gain achieved by PASS. Our results revealed that the array gain is maximized at an optimal number of pinching antennas and an optimal inter-antenna spacing. These findings underscore the importance of optimizing the two parameters for practical PASS deployments. Numerical results demonstrated that waveguide propagation loss has only a minor effect on the overall performance of PASS. However, the mechanical complexity associated with dynamically adding or removing dielectric materials remains an open issue and will be addressed in future research.
\bibliographystyle{IEEEtran}
\bibliography{mybib}
\end{document}